\documentclass[aps,prl,twocolumn,preprintnumbers,showpacs,amsmath,amssymb,nobalancelastpage]{revtex4}
\usepackage{amsfonts,mathrsfs,graphicx}
\usepackage{epsfig}
\usepackage{bm}
\def\be{\begin{equation}}
\def\ee{\end{equation}}
\def\bea{\begin{eqnarray}}
\def\eea{\end{eqnarray}}

\begin{document}
%%\preprint{ CERN-PH-TH/2011-132}

\title{Oscillations of the $^7$Be solar neutrinos inside the Earth}

 \author{A. N. Ioannisian$^{1,2}$, A. Yu. Smirnov$^{3,4}$, D. Wyler$^5$}
 \address{$^1$
Yerevan Physics Institute, Alikhanian Br.\ 2, 375036 Yerevan,
Armenia\\
$^2$Institute for Theoretical Physics and Modeling, 375036
Yerevan, Armenia\\
$^3$ Max-Planck Institute for Nuclear Physics, 
Saupfercheckweg 1, D-69117 Heidelberg, Germany\\ 
$^4$ICTP, Strada Costiera 11, 34014 Trieste, Italy\\
$^5$ Institut f\"ur Theoretische Physik,
 Universit\"at Z\"urich,
Winterthurerstrasse 190, CH-8057 Z\"urich, Switzerland }
%\ead{ara.ioannisyan@cern.ch}

\begin{abstract}
We explore in detail  oscillations of the solar $^7$Be neutrinos  
in the matter of the Earth.  
The depth of oscillations is about  $(0.1 - 0.2)\%$ and the length $\approx 30$ km.  
The period of the oscillatory modulations 
in the energy scale is comparable with the width of the line 
determined by the  temperature in the center of the Sun.  
The latter  means that depending on the length of trajectory (nadir angle) 
one obtains different degree of averaging of oscillations. 
Exploring these oscillations it is possible to measure the width of the $^7$Be line 
and therefore the  
temperature of the Sun,  determine precisely $\Delta m^2_{21}$,  
perform tomography of the Earth, in particular,  
measure the deviation of its form from sphere, and detect  small structures. 
Studies of the Be neutrinos open up a possibility to 
test quantum mechanics of neutrino oscillations and search for the sterile neutrinos.  
Accuracy of these measurements 
with future scintillator (or scintillator uploaded) detectors 
of the $\sim 100$ kton mass scale is estimated.  

\end{abstract}


\maketitle

%{\it Introduction.-}
\section{Introduction}
%%%%%%%%%%%%%%%%%%%%%%%%%%%%%%%%%%%%%%%%%%%%%%%%%%%%%%%%%%%%%%%%%%%%%%%%%%%

Exploration of the solar neutrinos moves to the phase of precision measurements.  
In future,  developments of the detection techniques and construction 
of the large mass detectors can open up various new possibilities
and one of them is  study of the solar $^7$Be neutrino oscillations 
in the Earth.  

The solar $^7$Be neutrinos have two salient properties: 
low energies  and  narrow 
width.  Due to environment effects in the Sun 
\cite{Bahcall:1993ej}  
the energy profile of the $^7$Be neutrinos,  $g(E)$,  is broadened 
and has an asymmetric form.  The maximum  value of $g(E)$,    
$E \sim$ 862.27 keV, is shifted to higher energy with respect to 
the laboratory value $E_{lab} \sim$ 861.64 keV  \cite{fn1}. 
The left (low energy) wing of the profile is
determined by the Doppler shift  caused by thermal velocities of
$^7$Be nuclei. The right (high energy) wing is mainly due to collisions determined 
by temperature at the center of the Sun (averaged over $^7$Be neutrinos production
region). The width of the line at the half of height equals
\be
\Gamma_{\rm Be} \simeq  1.6 ~{\rm keV},  
\ee
so that the relative size of the width is  
\be
\frac{\Gamma_{\! \! \rm Be}}{E} = 1.86 \cdot 10^{-3}. 
\ee

The width and the shift of maximum of $g(E)$ are proportional to  
the  central temperature of the Sun $T_c$.  
In \cite{Bahcall:1993ej} it was proposed to determine $T_c$  
by measuring the shift of the peak. 
It was  also mentioned that finite width of the $^7$Be spectrum 
affects  the depth of vacuum oscillations.

According to the standard solar model the flux of $^7$Be neutrinos is known with 
$1.4 \%$ accuracy.
For the LMA MSW values of oscillation parameters 
%%\cite{msw}
%%with oscillation parameters from the global fit 
the  $^7$Be neutrino  line is at 
the low energy edge of the so called transition region between the averaged 
vacuum oscillations and matter dominated conversion. The flux is suppressed 
mainly by the averaged vacuum oscillations probability with small additional suppression  
due to matter effect in the Sun.  
%%and no variations of the flux occurs apart from the 
%%seasonal variations related to eccenticity of the Earth orbit. 

The $^7$Be neutrino flux measured by BOREXINO~\cite{BOREXINOI}, 
\cite{Smirnov:2014hbh} and KamLAND \cite{kamland} 
is in a good  agreement with  the SSM predicted flux with 
suppression according the LMA solution. In fact, 
these measurements (reached $5\%$ accuracy)
provided important confirmation and consistency checks of the LMA MSW 
solution of the solar neutrino problem.  
%%as well as further tests of the solar neutrino model. 
Time dependence of the flux has been explored~\cite{Bellini:2011yj}. The data are in agreement 
with seasonal variations due to eccentricity of the Earth orbit and  no other variations 
have been found in agreement with LMA MSW expectations. 
In particular, after 2 years of exposure and 0.133 kton fiducial mass the following bound on 
the Day - Night asymmetry has been obtained \cite{Bellini:2011yj}: 
\be
A_{dn} = 2\frac{R_N - R_D}{R_N + R_D}  = 0.001 \pm 0.012 ({\rm stat}) 
\pm 0.007 ({\rm syst}). 
\label{adnbor}
\ee

On the way from the Sun  the coherence of neutrino state is lost,  
so that incoherent fluxes of the mass states arrive at the surface of the Earth. 
Inside the Earth these mass states  oscillate due to matter effect. 
According to  LMA MSW solution these oscillations 
proceed in the low matter density regime, and the expected effects
are very small (see e.g.  \cite{degouv}, \cite{jnb}  
\cite{ara1}, \cite{ara2}, \cite{ara3},  \cite{akhmedov}, \cite{aleshin}, \cite{aleshin2}).  
The effects are determined by the parameter:
\begin{equation}
\epsilon  \equiv  \frac{2 V_e E}{\Delta m^2_{21} }  \approx 
2.4\cdot10^{-3} \left( \frac{\rho}{2.7 {\rm {g \over cm^3}}} \right) \! \! 
                        \left( \frac{7.5 \cdot 10^{-5} {\rm eV}^2}{\Delta m_{21}^2}  \right) \! \! 
                        \left( \frac{Y_e} {0.5} \right), 
\label{eq:epsilon}
\end{equation}
where  $E$ is the  neutrino energy,  $V_e = \sqrt{2} G_F n_e $ 
is the matter potential  with $G_F$  and $n_e$ being the Fermi coupling constant and  
the electron number density correspondingly. 
The parameter  $\epsilon$ characterizes  deviations of the mixing angle
and the oscillation length in matter from their vacuum values: $\theta_{12}$ and  
\be 
l_\nu \equiv \frac{4\pi E}{\Delta m^2_{21}} \approx 28.5 {\rm km}
\left( \frac{7.5 \times 10^{-5}{\rm eV}^2}{\Delta m^2_{21}} \right).  
\ee
In fact, $\epsilon$ determines the depth of oscillations  of neutrinos with definite mass,  
$\approx 0.5 \epsilon$ which is about $\sim 0.1 \%$ in the mantle 
and $\sim 0.2\%$ in the core.     
Being of the order $10^{-3}$  the expected effects  are  
far beyond the present BOREXINO  (\ref{adnbor}) as well as expected SNO+ sensitivities.

Next generation of large (several tenth of ktons to hundred ktons) scintillator detectors 
like JUNO \cite{juno} or LENA~\cite{lena}  
will have sub-percent sensitivity to the  Day-Night asymmetry.  
Higher sensitivity can be achieved with 100 kton mass scale 
scintillator uploaded water detectors,  WBLS ~\cite{WBLS}. 
For 100 kton fiducial mass and 5 years exposure such a detector will collect 
$1.9 \cdot 10^3$ bigger statistics than the one used for the result 
(\ref{adnbor}). Correspondingly, the statistical error will be reduced 
down to $3 \cdot 10^{-4}$. So, if systematic 
errors is well controlled, the $0.1\%$ size 
Earth matter effects on the $^7$Be neutrinos 
can be established at about  $3\sigma$ level.  
 
%%
%%%%%%%%%%%ffff1%%%%%%%%%%%%%%%%%%%%%%%%%%%%%%%%%%%%%%%%%%%%%%%%%%%%%%%%%%
%%\begin{figure}[b]
%%\includegraphics[width=0.9\columnwidth]{width3.png}
%%\caption[...]{
%%The energy spectrum of the solar $^7$Be
%%neutrinos
%%(from \cite{Bahcall:1993ej}). Shown also is  the $\nu_e$ survival probability 
%%as function of the neutrino energy for several values of the nadir angle:   
%%$\eta = 1.4$ red dot-dashed line, $\eta = 1.4$ - green dashed line,  
%%$\eta = 0.6$ - blue line. 
%%
%%\label{Fig00}}
%%\end{figure}
%%%%%%%%%%%%%%%%%%%%%%%%%%%%%%%%%%%%%%%%%%%%%%%%
%%

%%In the present work we propose a new way to determine the Sun's
%%central temperature by looking at regeneration process of neutrinos in 
%%the Earth.

In this connection we will explore in detail the $^7$Be neutrino oscillations 
in the matter of the Earth. There are two very interesting coincidences 
related to the energy and width of the $^7 Be$ neutrino profile which  
allow one to obtain in principle unique information about neutrino properties, 
characteristics  of the $^7$Be neutrino spectrum,  properties of the 
Earth density profile and quantum mechanics of neutrino oscillations. 
We estimate possibilities of future large detectors to determine 
the width of the line, and consequently, the  central temperature of the Sun), 
to measure $\Delta m^2_{21}$ with unprecedent accuracy, to perform tomography of the Earth and 
search for very light sterile neutrinos.

The paper is organized as follows. In Sec. 2 we present 
relevant analytic results for the probabilities 
of oscillations in the Sun and the Earth 
as well as compute 
the relative variations of the flux with the nadir angle. 
In Sec. 3  the effects of averaging of the flux over the $^7$Be 
energy spectrum are explored. We compute the time (nadir angle) 
variation of number of events and estimate a potential  of 
future 100 kton scale detectors to establish the Earth matter effect, to  measure 
$\Gamma_{\rm Be}$  and $\Delta m^2_{21}$ in Sec. 4.  
Searches for sterile neutrinos are considered in Sec. 5. We conclude 
in Sec. 6.

\section{2. Oscillation in the Sun and the Earth}
%%%%%%%%%%%%%%%%%%%%%%%%%%%%%%%%%%%%%%%%%%%%%%%%%%%%%%%%%%%%%%%%

The electron neutrino $\nu_e$ produced in the center of the
Sun is adiabatically converted into the  combination
of the mass eigenstates $\nu_1$,  $\nu_2$, $\nu_3$. 
The combination  is determined by the mixing angles, $\theta^m_{12} = \theta_{12}^m (V_e^0)$, 
and $\theta_{13}^m (V_e^0)$  in the production point  
\begin{equation}
          \label{eq:sunpro}
\nu_e \ \rightarrow \ c_{13} \cos \bar{\theta}^m_{12}~\nu_1 \ +  
c_{13}\sin \bar{\theta}^m_{12}
~\nu_2  + s_{13} \nu_3,       
\end{equation}
where $c_{13} \equiv \cos \theta_{13}$,  $s_{13} \equiv \sin \theta_{13}$, 
and we neglected the matter effect on 1-3 mixing,   
so that $\theta_{13}^m (V_e^0)  \approx \theta_{13}$   
\footnote{Indeed, the influence of
matter on the 1-3 mixing is determined by  
$\epsilon (\Delta m^2_{31}) \sim 7 \cdot 10^{-5}$  and the observable effects are
further suppressed by smallness of $s_{13}$ and averaging of
oscillations associated with the third neutrino. For $\Delta
m^2_{31}$ the oscillation length is smaller than $1$ km. 
Interference between the modes of oscillations inside the Earth  driven by $\Delta
m^2_{31}$ and $\Delta m^2_{21}$ produces a negligible
effect. Thus we will use the vacuum value of 1 - 3 mixing.}.  
We use the standard parametrization of the PMNS mixing matrix.
In Eq. (\ref{eq:sunpro})  $\bar{\theta}^m_{12}$ is the value of 
1-2 angle in matter  averaged over the $^7$Be neutrino production region. 
It is given by  
\begin{equation} 
          \tan^2 2\bar{\theta}^m_{12} \ \approx  \ 
  \tan^2 2\theta_{12}  \left( 1-\frac{2c_{13}^2 \bar{V}_e^0 E}{\Delta m^2_{21} \cos 2 \theta_{12}} 
\right)^{-2},
\label{mixangle}
\end{equation}
where $\bar{V}_e^0$ is the average  matter potential  in the
production region.  

On the way from the Sun to the Earth the wave packets of mass eigenstates 
$\nu_i$ spread and separate.  At the production the lengths  of the wave packets 
in the configuration space equal 
\be
\sigma_x \approx \frac{2\pi}{\Gamma_{Be}} = 6 \cdot 10^{-8}\,  {\rm cm}.
\label{wp-width}
\ee
Due to spread their sizes become at the surface of the Earth 
\be
\sigma_x^{spread} =
\frac{m^2_i L_{sun} \Gamma_{Be} }{E^3},
\label{spread}
\ee
where $m_i$ is the absolute value of mass of $\nu_i$,  and
$L_{sun}$ is the distance from the Sun to the Earth.
For hierarchical spectrum, $m_2 \approx \sqrt{\Delta m^2_{21}}$, 
we obtain from (\ref{spread})  $\sigma_x^{spread} = 3.8 \cdot 10^{-6}$ cm,
({\it i.e.} 2 orders of magnitude larger than the original size of the packet). 
However,  separation of the wave packets of different mass eigenstates is larger:
$$
\Delta x  =  L_{sun} \Delta m_{21}^2/2 E^2 = 2 \cdot 10^{-3}~ {\rm cm}. 
$$
Even in the case of
degenerate spectrum, $m_1 \approx m_2 \approx 0.1$ eV, 
we obtain  $\sigma_x^{spread}/\Delta x \approx 0.25$. 
In addition there is also averaging of oscillations  
over the neutrino production region inside the Sun. 

Due to loss of coherence, neutrinos arrive at the surface of the
Earth as incoherent fluxes of $\nu_1$, $\nu_2$ and $\nu_3$
with relative admixtures given according to Eq. (\ref{eq:sunpro})  by
$c_{13}^2 \cos^2 \bar{\theta}^m_{12}$,  $c_{13}^2 \sin^2 \bar{\theta}^m_{12}$ 
and $s_{13}^2$ correspondingly. In matter of the Earth each 
of these mass states splits into eigenstates in matter and  oscillates. 

%%%%%%%%%%%%%%%%%%%%%%
%\begin{figure}[b]
%\includegraphics[width=\columnwidth]{1f.eps}
%\caption[...]{
%The dependence of $f(\Delta m^2,\theta)$ on
%$\tan^2 \theta$ for  $\Delta m^2 \ = \ 5 \times 10^{-5}$ eV$^2$, $
%7 \times 10^{-5}$ eV$^2$ and $10^{-4}  $eV$^2$
%\cite{Ioannisian:2002yj}.
%\label{Fig0}}
%\end{figure}
%%%%%%%%%%%%%%%%%%%%%%%%%%%%%%%%%%%%%%%%%%%%%%%%

%%Let us consider oscillations inside  the Earth.
The probability to find $\nu_e$ in the detector after crossing the Earth 
can be written as 
\begin{eqnarray}
\label{probnue}
P & = & c_{13}^2 (\cos^2 \bar{\theta}^m_{12} P_{1e}
  + \sin^2 \bar{\theta}^m_{12}  P_{2e}) + s_{13}^4 
\nonumber\\ 
 & = & 
  c_{13}^2 (\cos 2 \bar{\theta}^m_{12} P_{1e} + c_{13}^2 \sin^2 \bar{\theta}^m_{12}) 
+ s_{13}^4,
\end{eqnarray}
where $P_{1e}$ and $P_{2e}$ are the probabilities
of $\nu_1 \to \nu_e$ and  $\nu_2 \to \nu_e$
transitions in the Earth correspondingly.
Here we used  the unitarity relation: $P_{1e} + P_{2e} + s_{13}^2 = 1$ 
or $P_{1e} + P_{2e} = c_{13}^2$. 

The probability $P$ can be represented as
\begin{equation} 
\label{main1}
P  \equiv   P_D + \Delta P, 
\end{equation}
where
\begin{equation}
\label{main2}
P_D =  {c_{13}^4 \over 2 }(1  +  \cos 2 \bar{\theta}^m_{12}
\cos 2 \theta_{12} )  + s_{13}^4 \  
\end{equation}
is the probability during the day 
when $P_{1e} =  P^0_{1e}= c_{13}^2 \cos^2 \theta_{12}$, 
$P_{2e} = P^0_{2e}= c_{13}^2 \sin^2 \theta_{12}$,   
and 
\begin{equation} 
\label{main3}
\Delta P \  =  
c_{13}^2 \cos 2 \bar{\theta}^m_{12}~ (P_{1e} - P^0_{1e})   
\end{equation}
is the difference of the probabilities during the day and the night.    
The probability $P_{1e}$ equals~\cite{ara1}
 \begin{equation}
  \label{p1} 
P_{1e}/c_{13}^2 =
      \ \cos^2 \theta_{12} - {1 \over 2} \sin^2 2 \theta_{12}
     c_{13}^2 \int_{0}^{L} dx \  V_e(x) \sin \phi^m_{x \to L} ,
\end{equation}
where
\begin{equation}
\phi^m_{x \to L} (E) \equiv \int_{x}^{L} \! \! d x \ \Delta_m(x), 
\label{phasexl}
\end{equation}
\begin{equation}
\Delta_m(x) \equiv \frac{\Delta m^2_{21}}{2E} \sqrt{[\cos 2
\theta_{12} - c_{13}^2\epsilon(x)]^2 + \sin^2 2 \theta_{12}} \ 
\label{split}
\end{equation}
and $\epsilon$ is determined in (\ref{eq:epsilon}). 
Here $x$ is the distance between an  entrance point to the Earth and  a given point of trajectory. 

The oscillation length can be written as 
\be
l_m = \frac{2\pi}{\Delta_m(x)} = l_\nu [1 + \cos 2\theta_{12} c_{13}^2 \epsilon + O(\epsilon^2)].
\label{osclength}
\ee
The matter correction, $\cos 2\theta_{12} c_{13}^2 \epsilon$,  
is of the order $(1 - 2) \cdot 10^{-3}$. 

Notice that since in the Sun $P > 1/2$, or $\cos 2 \bar{\theta}^m_{12} > 0$,  
the  oscillations in the Earth  
suppress the survival probability,  $\Delta P < 0$,  in contrast to 
high energy Boron neutrinos for which $\cos 2 \bar{\theta}^m_{12} <  0$ 
and partial regeneration of the  $\nu_e$ flux occurs. 

The expression in  (\ref{p1}) is equivalent to  
the result of  adiabatic perturbation theory~\cite{ara2}:  
\begin{eqnarray}
\label{adiab} 
P_{1e}/c_{13}^2 &=&
      \ \cos^2 \theta_{12} - \sin 2 \theta^s_{12} 
\sin 2 (\theta^s_{12}-\theta_{12}) \sin^2 \phi_{0 \rightarrow L/2} 
\nonumber \\ 
&+&
\sin 2 (2 \theta^s_{12}-\theta_{12}) \! \! \!
      \int_{0}^{L}  \! \! \! d \theta_m(x) \  \cos \phi^m_{x \to L},
\end{eqnarray}
where $\theta^s_{12}$ is the mixing angle in matter  at the surface of the Earth. 
The angle $\theta^s_{12}$  is given by Eq.  (\ref{mixangle}) with $V_e$ substituted by 
$V_e (\rho_{surf})$. 
We use this formula for numerical computations inside different layers of the Earth. 

Combining Eqs. (\ref{main2}),  (\ref{main3}) and (\ref{p1}) we can write the relative 
variation of the $\nu_e-$flux  due to the Earth matter effect as 
\begin{equation}
 \label{releff}
A_e^0 \equiv \frac{\Delta P}{P_D}  = - \frac{c_{13}^2}{2} f(\Delta m^2_{21}, 
\theta_{12}, \theta_{13}) \int_{0}^{L} \!  dx \ V_e(x) \sin \phi^m_{x \to L},
\end{equation}
where
\begin{equation}
          \label{func}
          f(\Delta m^2_{21}, \theta_{12})  \ \equiv  \
          \frac{2 \cos 2\bar{\theta}^m_{12} \ \sin^2 2\theta_{12}}
          {1 \ + \ \cos 2 \bar{\theta}^m_{12} \cos 2 \theta_{12}  + 2 \tan^4\theta_{13}}.
          \end{equation}
For  $\theta_{12} = 34^{\circ}$  and $\Delta m^2_{12} =7.5 \times 10^{-5}$ eV$^2$ 
we obtain $f(\Delta m^2_{21}, \theta_{12}) \simeq 0.43$;  corrections due to  
the 1-3 mixing are below $0.1\%$. 
%%~\cite{Ioannisian:2002yj} [[what this reference means?]].
%%Last term in denominator f (\ref{func}) can be neglected. 

Quick estimation of the effect can be done  for constant density profile. 
In this case  
%%denoting by $\epsilon^c$ and  $\Delta_m^c$ the corresponding quantities 
we obtain $\phi^m_{x \to L} (E) =  \Delta_m^c (L - x)$, 
and in the lowest order in $\epsilon$ the Eq. (\ref{releff})  gives 
\be 
A_e^0 \approx - c_{13}^2 \epsilon  f \sin^2 \frac{1}{2}\Delta_m  L. 
\label{const}
\ee 
So, the depth of the oscillations  equals $ c_{13}^2 \epsilon f \approx \epsilon f$.

\begin{figure*}[t]
\includegraphics[width=\textwidth]{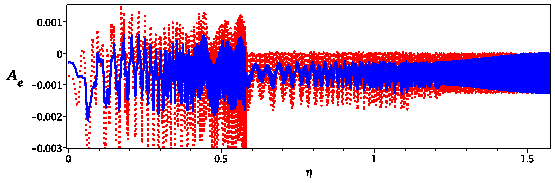}
\caption[...]{
The relative variations of the  electron
neutrino flux as function of the nadir angle of the neutrino trajectory. 
Dotted (red) line shows $A^0_e$ without averaging; 
solid (blue) line is $A_e$ which corresponds to the variations averaged
over the energy spectrum of the $^7$Be neutrinos. 
Spherical symmetry of the Earth is assumed. 
\label{Fig11}}
\end{figure*}

\begin{figure*}[!]
\begin{center}
\includegraphics[width=0.45\textwidth]{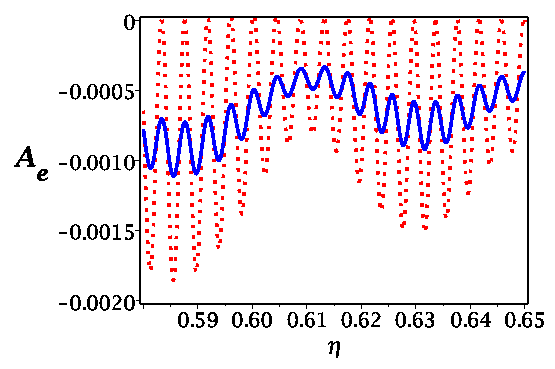} \hspace{0.1cm} %
\includegraphics[width=0.45\textwidth]{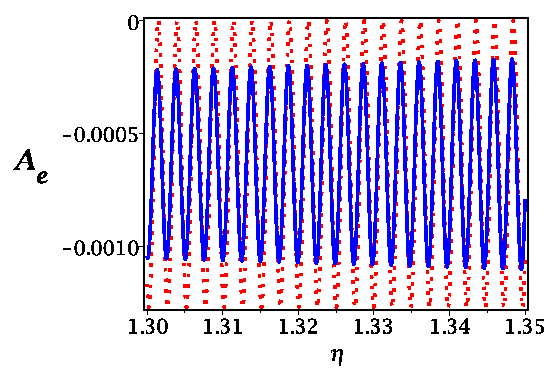}  %
\end{center}
\caption[...]{ The same as in Fig. \ref{Fig11}  with  two zoomed regions:
$\eta= [0.58 - 65]$ (left) and $[1.3 -  1.35]$ (right). 
\label{Fig33}}
\end{figure*}

Dependence of $\Delta P/P^0$ on the nadir angle $\eta$ is shown in Figs. \ref{Fig11},  
\ref{Fig33} (red lines). 
In our computations we used $P_{1e}$ from  Eq. (\ref{adiab}),  
the spherically symmetric Earth and the 5 layers parametrization 
of the density profile \cite{Lisi:1997yc}  
with sharp density jumps at 410 km,  660 km, 2830 km and 5150 km from 
the surface of the Earth.  
The length of the neutrino trajectory inside the Earth is given by 
$L = 2R_E \cos \eta$;  so dependence on $\cos \eta$ is equivalent to the dependence 
on $L$ and therefore Figs. \ref{Fig11} and \ref{Fig33} 
reflect dependence of oscillations on distance. 
Trajectories with $\cos \eta > 0.83$  cross the core of the Earth.  

According to the Figs. \ref{Fig11}, \ref{Fig33} (red lines) 
the depth of oscillations changes with the nadir angle which is related to 
breaking of adiabaticity. If the adiabaticity condition 
is fulfilled along whole the trajectory, the oscillation depth   
would be determined exclusively by the mixing angle at the surface of the Earth. 
(More precisely, by the mixing angle averaged over the distance of the order of 
oscillation length). This is satisfied  
for shallow trajectories with $\eta > 1.21$. At $\eta \approx  1.21$ neutrinos 
cross the first density jump  (410 km from the surface) and  
at $\eta \approx  1.11$ --  the second one  (660 km),  
where adiabaticity is broken,  and consequently, the depth of oscillations increases. 

The biggest change is at $\eta = 0.58$, where  neutrinos start to cross 
the core of the Earth. The depth of oscillations 
increases by factor of 2. 
At $\eta = 0.19$ neutrino trajectories 
cross also borders of the inner core. 
Breakdown  of the adiabaticity and   
oscillations in different layers lead also to modulations 
of the oscillatory picture which is related to the interference 
of effects from  different layers (see below).  

The length of the deepest
trajectory in the mantle  is $L_{max} = 10570$ km,   
therefore the number of periods in the mantle range, $\eta > 0.58$,  equals 
\be
\frac{x_{max}}{l_\nu} \approx 370.
\label{nperiods}
\ee
The number of periods  for the core crossing trajectories 
is bigger: about 420 for the vertical direction. 

%%%ffff2%%%%%%%%%%%%%%%%%%%%%%%%%%%%%%%%%%%%%%%%%%%%%%%%%%%%%%%%%%%%%%%
%%%%%%%%%%%%%%%%%%%%%%%%%%%%%%%%%%%%%%%%%%%%%%%%%%%%%%%%%%%%%%%%

%%%ffff3%%%%%%%%%%%%%%%%%%%%%%%%%%%%%%%%%%%%%%%%%%%%%%%%%%%%%%%%%%%%%%%
%%%%%%%%%%%%%%%%%%%%%%%%%%%%%%%%%%%%%%%%%%%%%%%%%%%%%%%%%%%%%%%%

%%%%%%%%%%%%%%%%%%%%%%%%%%%%%%%%%%%%%%%%%%%%%%%%%%%%%%
\section{3. Effect of the Be neutrino line width}
%%%%%%%%%%%%%%%%%%%%%%%%%%%%%%%%%%%%%%%%%%%%%%%%%%%%%%%%

\subsection{Averaging over the energy}  
%%%%%%%%%%%%%%%%%%%%%%%%%%%%%%%%%%%%%%%%%%%%%%%%%%%%%%%%%%%%%%%

Signal in the detector is determined by the variation, $A_e$, averaged 
over the energy spectrum of the $^7$Be neutrinos. Energy 
resolution of a  detector 
is much worse and therefore irrelevant for the oscillation picture. 
The $^7$Be neutrino flux can be written as 
\be
F_{Be} (E, t) = F_{Be}^0 (t) ~g(E), 
\ee
where $F_{Be}^0 (t)$ is the total flux 
and   $g(E)$ is the energy profile  normalized to 1: $\int dE g(E) = 1$.  
Dependence of $F_{Be}^0$ on time is due to 
the eccentricity of the Earth orbit around the Sun, as well as due to rotation of the  
Earth itself (due to bigger distance from the Sun at night).  
The former ($\pm 3 \%$) must be taken into account, the latter ($<10^{-4} \%$)  
can be neglected.

Variation of the $\nu_e-$flux averaged over  the $^7$Be  spectrum 
for the spherically symmetric Earth equals  
\be
\bar{A}_{e}(t) =   \int dE ~g(E) ~A_e (E, t).  
\label{appp}
\ee
The dependence of $\bar{A}_{e}$ on $\eta$ is shown in 
Figs. \ref{Fig11}, \ref{Fig33} (blue solid lines). 
In what follows we will present complete interpretation of the 
this dependence.  

1. {\it Suppression of the oscillation depth.} Integration over 
the energy profile leads to partial suppression of the oscillation depth 
which depends on the nadir angle. At the same time the integration does not affect 
the average value  of the  relative variations. 
The change of  the oscillation depth  with  the nadir angle 
is result of interesting and accidental coincidence:   
the width of the line, $\Gamma_{\rm Be}$,  is of the order of   
the period of oscillatory curve in the energy scale, 
$\Delta E_T$:  
\begin{equation}
 \Delta E_T \sim \Gamma_{\rm Be} 
\label{eq:cond}.
 \end{equation}
The period is defined by equality  
 \begin{equation}
 \frac{d \Phi_m}{dE} \Delta E_T  =  2 \pi,
\label{eq:period}
 \end{equation}
where
$\Phi_m \approx 2\pi L/l_m  \approx 2\pi L/l_\nu$ is the oscillation phase.  
We find from (\ref{eq:period})
\begin{equation}
\Delta E_T = E \frac{l_\nu}{L} = E \frac{l_\nu}{2 R_E \cos \eta}, 
\end{equation}
which depends on $\eta$. Therefore relation between  
$\Delta E_T$ and  $\Gamma_{\rm Be}$, and consequently, change of  
the degree of averaging and  the depth of oscillations 
depend on  $\eta$. For shallow trajectories (large $\eta$) 
$\Delta E_T \gg \Gamma_{\rm Be}$ the averaging effect is negligible.  
With decrease of   $\eta$  the period $\Delta E_T$ 
decreases and averaging becomes stronger  (see Fig. \ref{Fig11}). 
For the deepest trajectories in the mantle the suppression  is characterized 
by factor 2 - 3.  

In configuration space, partial averaging corresponds to
partial lost of coherence during propagation inside the Earth.
Indeed,  the relative shift of the wave packets of the eigenstates equals 
\be
\Delta x =  L \frac{\Delta m^2 }{2 E^2} =
5 \cdot 10^{-8} {\rm cm} \left(\frac{L}{10^4 {\rm km}}\right) =
0.8 \sigma_x \left(\frac{L}{10^4 {\rm km}}\right).  
\label{septtt}
\ee
In the last equality in (\ref{septtt}) we used estimation (\ref{wp-width}).  
The shift is of the order of original size of the wave packet,  
$\Delta x  \sim \sigma_x$,  which is complementary to the relation 
(\ref{eq:cond}). 

Since separation of wave packets is partial 
the effect of averaging
depends on the shape of the wave packet. So, 
measuring dependence of the depth of variations of signal on the 
base-line ($\eta$,   see Fig. 2)
one can in principle restore the shape of the wave packet.
(Loss of coherence 
is due to relative shift of the wave packets and the coherence condition 
is determined by the
original size of the wave packet and not by the length of the packet after spread \cite{kersten}).
This is very rare situation; another one is realized for 
supernova neutrinos propagating in the Earth \cite{kersten}. 

2. {\it Enhanced effect for the core crossing trajectories}. 
For trajectories  with  $\eta <  0.58$ the depth of oscillations 
averaged over the neutrino energy is factor of 4 - 5  
larger than for the deepest trajectories in the mantle. Furthermore, 
$\Delta P$ is positive during some part of the oscillation period. Both effects 
are due to higher matter density in the core and interplay of oscillation effect in the 
core and mantle. Appearance of the positive Earth matter effect,  $\Delta P > 0$,  
is also related to that fact that the average values  of $A_e$ in the core 
and in the mantle should be the same. 
The latter can be explained by the attenuation effect (see below): 
with strong averaging 
the core should not be seen. The only what one can observe is the average 
probability which is the same in the core and mantle.

3. {\it Attenuation effect.}  Integration over the energy  
leads to lost of sensitivity to the remote structures 
of the density profile~\cite{ara1}. This is characterized by the 
attenuation length, $\lambda_{att}$, in such a way that structures at distances 
from a detector bigger than $\lambda_{att}$ are not resolved in the oscillation picture.  
To find  the dependence of $\lambda_{att}$ on $\eta$ explicitly  we  take  
for $g(E)$  a Gaussian profile  
with the width  $\Gamma_{\rm Be}$.   
(The use of exact line shape does not change this result 
substantially.)  In this case the expression (\ref{appp}) 
gives
\begin{equation}
{A}_e = -  {1 \over 2} c_{13}^2 f(\Delta m^2, \theta)
      \int_{0}^{L} \!   dx \ V(x) F(L-x) \sin \phi^m_{x \to
      x_f},
\label{eq:ae-av}
\end{equation}
where
\begin{equation}
 F(L-x)= {\rm exp} \left\{-2 \left[ {\pi \Gamma_{\rm Be}  (L-x) \over E \ 
l_\nu }\right]^2 \right\}. 
\label{FF}
 \end{equation}
According to Eqs. (\ref{eq:ae-av}) and (\ref{FF}) a  contribution 
to the integral (\ref{eq:ae-av}) from  structures situated at the distance 
$(L-x) > \lambda_{att}$ from detector, where the attenuation length 
$\lambda_{att}$ is determined by the relation 
$$
\sqrt{2} \frac{\pi \Gamma_{\rm Be}  \lambda_{att}}{E  l_\nu }   =  1,   
$$ 
becomes exponentially suppressed.   
This gives  
$\lambda_{att} \sim  7000$ km which is comparable with radius of the Earth and bigger than the 
distance to the core. 
So, the core of the Earth 
should be seen in agreement with result of   Fig. \ref{Fig11} 
(increase of the oscillation amplitude at $\eta = 0.58 $).    
The  average values of  oscillation curve in the mantle and the core 
are nearly the same.

4. {\it Modulation of the oscillation picture; interference. } 
At $\eta =  1.21$ and $\eta =  1.11$   
the oscillatory picture  changes:  the depths of oscillations slightly increases, and 
high frequency oscillations start to be modulated by low frequency modes. 
Periods of modulations increase with decrease of $\eta$ (see Fig. \ref{Fig11}).  
The modulations originate from  the  density  
jumps at  distances $h_1 = 410$ km  (by $5\%$) and   
at $h_2 = 660$ km (by $10\%$) from the surface of the Earth. 
These density changes occur over depth interval less than 
5 km \cite{shearer}, \cite{petersen1993},        
which is much smaller than the oscillation length. Therefore  
the  jumps break adiabaticity.  The adiabaticity is well satisfied 
within the layers  bounded by the jumps. 
The jumps lead to increase  of 
the depth of oscillations at $\sin \eta = 1 - h_i/R_E$. 
The periods of modulations  in the $\eta$ scale 
can be obtained in the following way. 
The length of the trajectory with a given 
value of the nadir angle $\eta$ between the surface of the Earth and 
the $i$th jump  equals 
\be
x_i(\eta) = R_E \cos \eta - \sqrt{R_E^2 \cos^2 \eta - 2 R_E h_i + h_i^2}. 
\ee  
Consequently, the length of trajectory in the layer between the i-th and j-th jumps is 
$x_{ij} = x_i(\eta) - x_j(\eta)$. 
Then the period of modulations, $\delta (\eta) = \eta_1 - \eta_2$ 
is determined by the condition 
\be
[x_i(\eta_1) - x_j(\eta_1)] - [x_i(\eta_2) - x_j(\eta_2)] = l_m.  
\label{period}
\ee
In the approximation of flat layers ($R_E \rightarrow \infty$) or 
$R_E \cos \eta \gg h_i$ the condition simplifies: 
the length of the trajectory  equals 
approximately $h_i/\cos \eta$ and instead of  
(\ref{period}) we obtain 
\be
\Delta (\cos \eta) =  (\cos \eta_1 - \cos \eta_2) 
\approx \frac{l_\nu}{h_i} \cos^2 \eta. 
\label{deltt}
\ee
From (\ref{period}, \ref{deltt}) for $\eta = 0.58$ 
(the deepest mantle trajectory) we find $\Delta \eta = 0.08$ in agreement 
with the results of Fig.~\ref{Fig11}. With increase of $\eta$ the period of 
modulations decreases. 
Thus, characteristics of modulations  encode information about parameters of non-adiabatic 
density jumps (their position and size). 
Modulations become especially profound for the deepest  mantle trajectories. 

\subsection{Geophysics of the Be neutrino oscillations} 
%%%%%%%%%%%%%%%%%%%%%%%%%%%%%%%%%%%%%%%%%%%%%%%%%%%%%%%%%

At the sub-percent level of the experimental accuracy 
and very short oscillation length 
a number of new effects become accessible and should be taken into account. 
Recall that the oscillatory curves shown in Figs. \ref{Fig11}, \ref{Fig33}
correspond to the ideal spherically symmetric Earth. In reality the Earth profile
is not symmetric and deviations are of the order of 
 oscillation length. Deviations include:

(i) Non-sphericity of the Earth as whole.

(ii) Small scale structures at the surface (mountains, oceans, {\it etc.}). 

(iii) Structures in the crust (layers with anomalous density, cavities). 

(iv) Possible structures in the mantle and the core, in particular  deviation of the core 
from spherical form. 

Effects of the deviations on $A_N$ can be quantified by  contributions
to the total oscillation phase $\Phi_m$ which can be written 
according to (\ref{phasexl}), (\ref{split}) as 
\bea
\Phi_m & = & 2\pi \frac{L}{l_\nu} \left[1 - \cos 2\theta_{12} c_{13}^2 
 \frac{1}{L} \int dx \epsilon(x) \right]
\nonumber\\
&  = &  
2\pi \frac{L}{l_\nu} (1 - \cos 2\theta_{12} c_{13}^2  \bar{\epsilon}).   
\label{phasetot}
\eea
Here $\bar{\epsilon}(L)$ is the averaged value of the matter  
parameter over the neutrino trajectory with the length $L$.  
The matter correction  to $\Phi_m$ (last term in (\ref{phasetot})) equals  
\be
\Delta \Phi_m = 2\pi \frac{L}{l_\nu} \cos 2\theta_{12}  \bar{\epsilon} 
= \frac{\pi}{2}\left( \frac{\bar{\epsilon}}{10^{-3}}\right)
\left( \frac{L}{7000 {\rm km}} \right) .  
\label{phasematt}
\ee
For $\bar{\eta} = 1.2$ (the outer mantle) 
we obtain $\Delta \Phi_m = 0.9 \pi$. 
For $\bar{\eta} = 0.58$ (the deepest trajectory in the mantle)  
the corresponding number is  
$\Delta \Phi_m = 3\pi $. For 
$\bar{\eta} = 0.4$ (the core crossing trajectory) 
we find  $\Delta \Phi_m = 5.9 \pi $. 
So, to avoid an uncertainty in the phase, $\bar{\epsilon}$ 
should be known with  accuracy better than  
$20\%$, $8\%$ and $4\%$ in the outer mantle, deep mantle  and core crossing trajectories.   
In general, all possible corrections and uncertainties (e.g. in $L$, $\Delta m^2_{21}$, 
{\it etc.}) which  change the phase by $\sim \pi/6$ should be taken into account. 
In view of many oscillation periods obtained over the baseline even small 
effects may become important. 
Large uncertainties in  the density profile would wash out 
the oscillatory dependence.    

Let us consider some  effects of density profile perturbations.  

1. Deviation of the Earth shape from ideal sphere can be characterized as follows. 
The polar and equatorial radii 
equal  $R_{pol} = 6356.7$ km and  $R_{eq} =  6378.1$ km correspondingly. The   
difference in diameters is about 43 km.
For a given  latitude $\psi$ the distance from the surface to the center 
of the Earth can be approximated by 
$$
R_\psi \approx R_{eq} - (R_{eq} - R_{pol}) \cdot \sin^2 \psi  
= R_{eq} - 21.5 {\rm km} \cdot \sin^2 \psi. 
$$   
Introducing the average radius of the Earth as
$\bar{R} = 0.5(R_{eq} + R_{pol})$ we obtain  the deviation
from the average 
$R_{\psi} - \bar{R} = (R_{eq} - R_{pol})(0.5 - \sin^2 \psi)$. 
E.g. for equator ($\psi = 0$) the difference of   trajectory
lengths can be as large as $(R_{eq} - R_{pol})/2 =  21.4$ km.

Due to non-sphericity of the Earth  
for the same value of the nadir angle the length of trajectory
and consequently the oscillation phase depend on the azimuthal angle.
Therefore deviation from sphericity leads to modification  of the oscillatory
curves in Figs. \ref{Fig11}, \ref{Fig33}. 
If  variable $\eta$ is used,  one needs
to average phase over the azimuthal angle. This however, may lead to
complete averaging of the oscillations.
Indeed, for a detector at the  latitude $\psi$ and for small nadir angle $\eta$ 
the difference of lengths of trajectories
for different azimuthal  angles  can be as large as  
\be
\delta x \approx 2 \bar{R} (r - 1) \tan \eta \sin 2 \psi 
\ee
It  can be about  (10 - 20) km  even for small $\eta$. 

Inversely, due to non-sphericity the trajectories with the same $L$ 
have different $\eta$ and azimuthal angles and therefore 
$L$ does not fix the density profile uniquely.  
So $\bar{\epsilon}$ can be different 
for trajectories with the same $L$ but different $\eta$. This variation,  however, 
can be neglected in the first approximation, and  one can use $L$ as 
the parameter to mark events.

2. The effect of small structures at the surface of the Earth  
depends on $\eta$.
Let us neglect here non-sphericity and use $\eta$ variable.  
Small structures  produce distortion  
of the periodic sinusoidal curves shown 
in Figs. \ref{Fig11}, \ref{Fig33} 
 leading to  appearance of irregular perturbations, 
substructures, shifts of maxima and minima, 
{\it etc.}.

Consider perturbation produced by a structure 
at the point where neutrino enters the Earth (the beginning of the trajectory) 
which has the surface length $x_{pert}$ and height $h_{pert}$. 
Modification  of  $A_e(\eta)$ produced by this structure 
depends on relative values of  $x_{pert}$ and the distance $x_T$ which corresponds to 
$\eta_T$ --  period of $A_e(\eta)$ in $\eta$.  
In turn, $\eta_T$ is determined from the  equation 
\be
2R_E \left[\cos \eta - \cos (\eta + \eta_T)\right] = l_m \approx l_\nu   
\ee 
which can be rewritten as 
\be 
\sin \eta \sin \eta_T  +  \cos \eta (1 -  \cos \eta_T) = \frac{l_m}{2R_E}.    
\label{etat}
\ee 
Then the distance $x_T$ as function of  $\eta$ is given by 
\be
x_T \approx \sqrt{l_\nu^2 + [4R_E \cos \eta \sin (0.5 \eta_T)]^2}.  
\label{xperiod}
\ee

For very  shallow trajectories ($\eta \sim 1.5$),  
$x_T \sim l_\nu \sim 28$ km,  which is comparable with the length of 
perturbations. 

With decrease of $\eta$ the $x_T$ increases. 
For directions not very close to vertical the first term 
in (\ref{etat})  dominates and we obtain 
$\sin \eta_T \approx l_m/(2R_E \sin \eta)$, or 
\be
\eta_T \approx  \frac{l_m}{2R_E \sin \eta}. 
\label{etatt}
\ee 
Insertion of  this $\eta_T$ into  (\ref{xperiod}) gives 
\be
x_T \approx \frac{l_\nu}{\sin \eta}. 
\ee 
E.g. for the deepest trajectories in the mantle, $\sin \eta = 0.56$,  
we obtain $x_T = 50$ km. 
$x_T$ increases faster near vertical directions. 
For $\eta = 0$ we find from (\ref{etat}) and (\ref{xperiod})
$$
x_T \approx 2 \sqrt{l_\nu R_E} \approx 850 ~{\rm km}.  
$$

The size of perturbation of  $A_e$  is given by the additional 
phase acquired due to perturbation of the profile:   
$$
\Phi_{pert} \approx  2 \pi  \frac{L_{pert}}{l_\nu },  
$$
where $L_{pert} \approx h_{pert}$ for  nearly vertical trajectories,  and 
$L_{pert} \approx x_{pert}$ for nearly  horizontal trajectories. 
The change of $A_e$ equals
\be
\Delta A_e \approx \frac{1}{2} c_{13}^2 \epsilon f [\cos \Phi_m  - \cos (\Phi_m + \Phi_{pert})]. 
\ee

For shallow trajectories already $l_{pert} \approx x_{pert} = 5$ km become important. 
$\Phi_{pert} = 0.35 \pi$ and it will change $A_e$ in the interval 
$\Delta \eta \sim x_{pert}/ x_T \approx x_{pert}/l_\nu = 0.17$ of the period.  
The  changes of $A_e$ in units $A_e^{max} = -c_{13}^2 \epsilon f$ 
are $0 \rightarrow 0.2$, $0.5 \rightarrow 0.95$, $1 \rightarrow 0.73$ for 
$\Phi_m = 0, ~\pi/2, ~\pi$  correspondingly.  This also means that maximum 
and minimum will be shifted. For $x_{pert} = 15$ km the effect will 
be of the order 1, e.g. maximum will become minimum  and vice versa. 
In this case the change can be viewed as a local shift of the oscillatory curve 
by half of period in $\eta$. 

With decrease of $\eta$ effect of  perturbation of the same size  will decrease since 
$x_T$ becomes larger and $l_{pert}$ becomes smaller. That is,  the change will be over 
smaller part of the period and the change of $A_e$ will be smaller.  
 
For vertical trajectories we have $x_{pert} \ll x_T$ and 
$\Phi_{pert} \leq 0.35 \pi$. So, the effect shows up as small  
perturbations of oscillatory curve with typical size being much smaller than the period.

Since the oscillation length is about 30 km, small structures at the surface of the Earth 
(mountains, oceans, seas) as well as  in the 
crust may become important. Relatively small structures can be  
averaged out when  integrating over  the nadir angle intervals.

%%%ffff3%%%%%%%%%%%%%%%%%%%%%%%%%%%%%%%%%%%%%%%%%%%%%%%%%%%%%%%%%%%%%%%
\begin{figure*}[!]
\begin{center}
\includegraphics[width=0.45\textwidth]{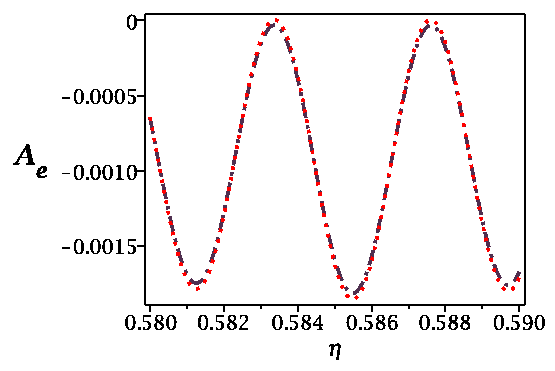} \hspace{0.1cm} %
\includegraphics[width=0.45\textwidth]{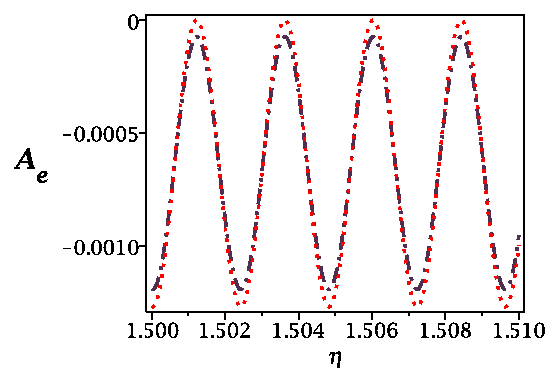}  %
\end{center}
\caption[...]{Effect of averaging over the production region
of $^7$Be neutrinos in the Sun. 
The relative change of the electron
neutrino flux for mantle crossing  trajectories with   $\eta=0.58
\dots 0.59$ (left) and $1.5 \dots 1.51$ (right)  without (dotted line)
and  with (dash-dotted line) averaging.   
\label{Fig34}}
\end{figure*}
%%%%%%%%%%%%%%%%%%%%%%%%%%%%%%%%%%%%%%%%%%%%%%%%%%%%%%%%%%%%%%%%

3. It is more difficult to control underground structures. 
Although  variations of density in regions of (5 - 30) km size  
do not produce significant change of the phase and their effect 
will be subleading~ \cite{ara1}.  In principle one can study the  
shape of the core of the Earth and its deviation from sphere.\\

In this connection let us estimate effect of 
finite size of the $^7$Be neutrino production in the Sun. 
From the Earth the  $^7$Be neutrino production region in the Sun is seen as a disk of the 
angular size $\eta_{disc}\sim 10^{-4}$ radians. 
Neutrinos produced in different parts 
of the disc will have slightly different 
nadir angles and therefore baselines. Therefore one needs to perform integration 
over the corresponding intervals of the nadir angle. 
The value of  $\eta_{disc}$ is an order of magnitude smaller 
than  the oscillatory period  $\Delta \eta \approx 0.002 - 0.003$, 
so the effect of averaging over disc is expected to be small.  

To quantify this  we approximate the distribution of the neutrino sources in the Sun  
by 
\begin{equation}
f(\eta^\prime, \eta) = {1\over \delta_\eta \sqrt{2\pi}} \ 
e ^{-{(\eta-\eta^\prime)^2 \over 2 \delta^2_\eta}}
\end{equation}
with  $\delta_\eta \simeq 1.9 \times 10^{-4}$. 
The result shown  in  Fig.~\ref{Fig34}, indeed, confirms the smallness of the effect. 
Averaging over the disc is slightly stronger for large nadir angles 
where the oscillatory period is smaller. \\

\section{4. Observational effects}
%%%%%%%%%%%%%%%%%%%%%%%%%%%%%%%%%%%%%%%%%%%%%%%%%%%%%%%

Observable effects consist of certain time variations
of the signal during the nights, $A_N(t)$.
Variations have quasiperiodic dependence on time,
the  amplitude of variations also  changes 
during the night, furthermore oscillations with high frequency
are modulated by lower frequencies. 
Sinusoidal dependence is modified by non-sphericity of the Earth, small structures of the 
profile, etc.. Parameters of these variations
depend on position of a detector.

In our estimations of sensitivities to different effects 
we will use for simplicity the spherical Earth.
Deviations from sphericity should be taken into account
for different detectors individually.
As function of $L$ the oscillatory curve has regular dependence
which is similar to the dependence on  $\cos \eta$ for spherical case.
Local deviations from sphericity produce complicated effects and 
perturbations of the profile should be  
taken into account. 
%%exact density profile will not change our estimations of sensitivity 
%%significantly. 

In the case of spherical Earth and without small structures  
we can use $z \equiv \cos \eta $  as variable and evaluate 
number of events in small $\Delta z$ bins.  
Recall that different detectors cover the nadir angle ranges  
with minimal  $\eta_{min}$  which depends on the latitude of the detector.

\subsection{Rate of events in the detectors}
%%%%%%%%%%%%%%%%%%%%%%%%%%%%%%%%%%%%%%%%%%%%%%%%%%%%%%%%%%%%%%%

The rate of $\nu - e$ scattering  
events in a  detector can be written as 
\be
R(t) = N_e \int dE F_{Be} (E, t) \sigma_e (E) [P(E, t) + r_{NC}(1 - P(E, t))], 
\label{ratett}
\ee
where $N_e$ is the total number of electrons in a fiducial volume,   
$\sigma_e$ is the cross-section of $\nu_e - e$ scattering,  
$r_{NC} \equiv \sigma_{NC}/ \sigma_e$,  and $\sigma_{NC}$ is the 
cross-section of the $\nu_\mu - e$ scattering due to neutral currents.  
The last term in (\ref{ratett}) accounts for  the contribution 
of the $\nu_\mu$ and $\nu_\tau$  
neutrinos that are generated by oscillations. 
In the spherical Earth approximation dependence of the probability on time is 
via $\eta(t)$ only: $P(E, t) = P(E, \eta(t))$ 
%%In reality it also depends on the azimuthal angle: $P(E, t) = 
%%P(E,\eta(t), \phi (t))$.  
The rate  (\ref{ratett}) can be rewritten as 
\be
R_N(t) = N_e F_{Be}^0 (t) \bar{\sigma}_e \int dE ~g(E)~ [ P(E, t)(1 - r_{NC}) + r_{NC}],  
\label{rate}
\ee
where $\bar{\sigma}$ is the averaged (over the $^7$Be line) value of cross-section.    
 
Let us  introduce the relative variations of the rate of events due to 
the Earth matter effect  
\be
A_{N}(t) = \frac{R_N(t) - R_D(t)}{R_D(t)}.  
\label{ant}
\ee
Here $R_D(t)$ is the rate which would be without Earth matter effect and it  
is given by Eq.  (\ref{rate}) with $P_N$ substituted by $P_D$. 
Inserting (\ref{rate}) into  Eq.  (\ref{ant}) and assuming that $\theta_{12}^m$,  
and consequently,  $P_D$  do not change with energy in the interval  
of energies of the  $^7$Be neutrinos,  we obtain
\be 
A_{N}(t) = A_e(t) \kappa,  
\label{anevent}
\ee
where 
\be
\kappa \equiv  \frac{1 - r_{NC}}{1 + r_{NC}(1/P_D - 1)} 
\ee
gives the correction due to contribution of  $\nu_\mu$  and $\nu_\tau$. 
For  $r_{NC} = 0.2$  and $P_D \approx 0.6$ we obtain  $\kappa = 0.7$.     
Thus, the correction leads to damping of variations of signal.

For estimations we will consider future 
scintillator (or scintillator uploaded) detector  
with fiducial mass $M_D = 100$ kton and 5 years exposure  as illustrative values. 
By simple re-scaling one can find effects in specific proposed detectors such as 
LENA, JUNO or WBLS.  

To evaluate the total number of events 
in such a detector we will use the rate estimated 
%%in  \cite{Mollenberg:2014mfa}: 
%%$1.7 \cdot  10^4$ events per day in 35 kt fiducial mass and 
in \cite{Wurm:2010mq} (see also \cite{Mollenberg:2014mfa}):  
$1.5 \cdot 10^4$ events per day in 48 kt. 
%%[[do these estimations take into account effect of oscillations in the Sun and 
%%scattering due to NC?]] 
Normalizing to this number 
%%for 100 kton during 5 years 
we  obtain 
\be
N^{tot} = 
%%(5.7 - 8.9) 10^{7} 
5.7 \cdot 10^{7} 
\left(\frac{M_D}{ 100{\rm kton}}\right) 
\left(\frac{t}{ 5 {\rm years}}\right)~  {\rm events}. 
\label{totnev}
\ee
These events can be analyzed in various  
ways to make the data sensitive to different quantities.

\subsection{Establishing the Earth matter effect}
%%%%%%%%%%%%%%%%%%%%%%%%%%%%%%%%%%%%%%%%%%%%%%%%%

For this it is enough to determine  value  $A_{N}$
averaged over $\eta$. 
So, one should compare the total numbers of events detected during the 
nights and  days during whole the exposure. 
(Possible distortion of $A_N(t)$ by perturbations of the density profile 
is not relevant here.) 
We take that approximately half of the total number of 
events (\ref{totnev}) is detected 
during nights and another half during days: 
\be
N_{N} \approx N_D = 
%%(2.8 - 4.5) 
2.85 \cdot 10^{7} ~ {\rm events}.  
\ee 
Then difference of the numbers of night and day events 
due to oscillations in the Earth  equals
\be
N_{D} - N_N \approx - N_D {A}_N = 
- N_D \kappa A_e  = 
%%- (1.7 - 2.2) 
 1.7 \cdot 10^{4} ~ {\rm events}. 
\ee 
The statistical error of  measurements of $N_D - N_{N}$ is 
$\sigma_N  =   \sqrt{2N_N} = 
%%(7.5 - 9.5) 
7.5 \cdot 10^{3}$.  
So, after 5 years 
\be
(N_D - N_{N}) \approx 2.3 ~\sigma_N. 
\ee
That is,  the difference of the night and day signals 
can be established at $2.3 ~\sigma$ level. 
This is in agreement with estimation made in the introduction. 
Notice that various systematic uncertainties cancel in the 
relative variations.

\subsection{Variations of the signal during nights}
%%%%%%%%%%%%%%%%%%%%%%%%%%%%%%%%%%%%%%%%%%%%%%%%%%%%%%%%

One can measure  oscillatory variations of the $^7$Be neutrino signal 
during nights (Fig.~\ref{Fig11}) 
detecting events in short  time intervals. 
This method may have less systematics than the first one described in the previous subsection. 
However here new systematics may appear due to unaccounted 
effects of small structures of the Earth 
(see below). Period of time variations 
can be estimated for spherically symmetric profile in the following way.  
The period in the nadir angle scale is given in (\ref{etatt}), $\eta_T = 2.24 \cdot 10^{-3}/\sin \eta$. 
Then the period in time equals  
\be
t_T = \frac{l_m}{ 2 R_E \sin \eta} \left(\frac{d \eta}{d t }\right)^{-1},  
\ee
where the speed of change of the $\eta$ with time  depends on value of  $\eta$.    
%%and can be estimated as 
%%\be
%%\left(\frac{d \eta}{d t }\right)  \approx (0.15 - 0.26) \frac{rad}{hour}. 
%%\ee
We find that the average period  equals about 1 min.  
So,  to measure  the oscillatory curve one needs to take time intervals smaller than 15 sec. 
Number of events expected in such an interval  will be about 3 - 5. 
Therefore summation of  signals from  time intervals in which the Earth 
matter effect is the same during several years is needed.  

In the approximation  of spherically symmetric  
Earth  the length of trajectory and the density 
profile are  fixed uniquely by the nadir angle $\eta$, and the Earth matter 
effect is quasi-periodic  function of  $\cos \eta$ (since $L = 2R_E \cos \eta$).  
Then from  (\ref{phasetot}) we obtain the period in $\cos \eta$ 
\be
(\cos \eta)_T = \frac{l_\nu}{2 R_E 
(1 - \cos 2\theta_{12} c_{13}^2  \bar{\epsilon})}. 
\label{deltatt}
\ee
Due to increase of $\epsilon(\eta)$ with decrease of $\eta$ 
the period will slightly increase for deeper trajectories.   
The relative change of period is of the order $10^{-3}$. 
It would be exactly periodic function in the case of constant density.  

So,  the method consists of splitting the whole  $\cos \eta$ interval 
for a given detector into small intervals $\Delta (\cos \eta) <  (\cos \eta)_T$,  
identification  the corresponding time intervals during nights,  
$\Delta t = \Delta t (\cos \eta,  \Delta (\cos \eta))$,  
and accumulation of  events in these small intervals during several years.

However, even with 100 kton detector it is not possible to measure whole  
the oscillatory curve of Fig. \ref{Fig11} with appreciable statistical significance. 
Therefore one should also sum up the signal over all periods of Fig. \ref{Fig11}. 
Introducing corrections due to  change of period with $\eta$ one can 
combine all the events in a single  period using the 
effective phase (\ref{phasetot}) as variable. 
This means that for each event or several events detected during small 
enough time intervals $\Delta t = (10 - 15)$ sec, one finds   
$\Phi_m (t)$, and then collects  events for the intervals $\Phi_m (t) + 2\pi k$. 
%%(In any case one should average over production region in the Sun. )

Following this we divide  all the events detected during the nights in to 
two groups: events  detected in the first half of  period, $N_1$,  
and events detected in the second half  of  period,  $N_2$. 
Then  difference of the events equals  
\be 
N_1 - N_2 = \langle D \rangle 
N_N \approx   \langle D \rangle  N_D,  
\ee
where 
\be
\langle D \rangle =  \frac{2}{\pi} (A_N^{max} - A_N^{min}).  
\label{avdiff}
\ee
Here  $A_N^{max}$ and  $A_N^{min}$ maximal and minimal values of 
$A_N$ (averaged over different periods) and 
factor $2/\pi$ reflects decrease of the difference due to integration 
over half  periods as compared to the total depth.  Thus, $\langle D \rangle$ 
is the relative variation of number of events integrated over half a period.  
From Fig. \ref{Fig11} we find that for the mantle trajectories 
$  \kappa (A_N^{max} - A_N^{min}) \approx 5.2 \cdot 10^{-4}$. 
Taking  $N_N  \approx 
%%(2.8 - 4.5) 
2.8 
\cdot 10^{7}$, we obtain 
$N_1 - N_2 = 0.9 \cdot 10^{4}$. 
The statistical error of measurements of the  
difference $(N_1 - N_2)$ is $ \sqrt N_N = 5.3 
\cdot 10^{3}$. Correspondingly,  
variations during the night can be established at 
$1.8 \sigma$ level. 
  
Let us estimate effect from the core crossing trajectories. 
We assume that fraction of events collected from these trajectories 
is about $10\%$, that is,  
$N_{core} = 
%%(2.8 - 4.5)
2.8 \cdot 10^{6}$.   
The depth of modulation is larger:  $A_N^{max} - A_N^{min} = 2 \cdot 10^{-3}$. 
The expected difference of events in the first and  second halves of the period 
equals $N_1 - N_2 = 
%%(2.5  - 4.0) 
2.5 
\cdot 10^{3}$. 
The statistical errors is  
$\sigma = \sqrt{N_D} = 
1.7 
%%(1.7 - 2.1) 
\cdot 10^3$. 
Therefore  $N_1 - N_2 = 1.5 
%%(1.5  - 1.9) 
\sigma$. 
Summing up the significances in the core and  mantle  we obtain that 
variation of the signal can be established at 
$ 2.3 \sigma$ level. These estimations have been performed for the ideal periodic oscillatory 
dependence of $A_N$ on $\cos \eta$. As we discussed in Sec. 3, 
the presence of small structures  at the surface of the Earth and in the crust 
distorts the sinusoidal dependence, in particular it can shift significantly maxima and minima 
in the $\cos \eta$ scale.  
This leads to uncertainties in determination of borders of periods and intervals 
in which events should be summed up. 
The uncertainties can be reduced if a profile of the Earth along trajectory of each event is 
known and therefore for each event the oscillation phase can be determined. 
Actually, it will be enough to know the length of  trajectory, and the rest can be accounted
as systematic error. As a result, significance of establishing of variations 
may  be somehow lower.

%%%%%%%%%%%%%%%%%%%%%%%%%%%%%%%%%%%%%%%%%%%%%%%%%%%%%%%%%%%%
\subsection{Determination of the line width}
%%%%%%%%%%%%%%%%%%%%%%%%%%%%%%%%%%%%%%%%%%%%%%%%%%%%%%%%%%

%%%%%%%%%%%%%%ffff5%%%%%%%%
\begin{figure}[t]
\includegraphics[width=\columnwidth]{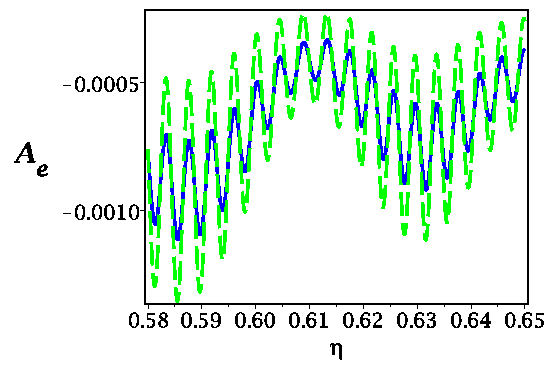}
\caption[...]{
The relative change of the electron 
neutrino flux for the mantle crossing  trajectories   
as the function of $\eta$  for two different values of width of the 
$^7$Be line which correspond to two different temperatures in the center of the Sun: 
%%$\eta= 0.58 \dots 0.65$. 
Solid line for sun central temperature $T_\odot =15.5 5\times 10^6$ (solid line)
$7.77\times 10^6$ degree (dashed line).
\label{fig:temp}}
\end{figure}
%%%%%%%%%%%%%%%%%%%%%%%%%%%%%%%%%%%%%%%%%%%%%%%%

The depth of oscillations in the Earth  decreases  
with increase of the width of the line which in turn is proportional to 
the central temperature in the Sun:   
$\Gamma_{\rm Be} \propto T_c$. 
In Fig. \ref{fig:temp} we show the oscillatory picture for two different 
widths of the line which correspond to two different central temperatures of the Sun.     
As follows from  Fig. \ref{fig:temp},   the depth of oscillations increases by 
factor 2 with decrease of the temperature by the same factor. 
 
According to Fig.  \ref{Fig11} for shallow trajectories, $\eta > 1.4$,  the averaging effect is 
negligible and the depth of oscillation is the same as for monochromatic spectrum. 
The averaging effect strengthens and depth decreases with decrease of $\eta$. 
For larger width $\Gamma_{\rm Be}$, the   change of  depth is faster. 
The depth of oscillations is larger for the core-crossing trajectories and  
it changes 
with  $\eta$ weakly. Therefore  methods of determination of the width can be  based on 
determination  of the depth of variations $A_N$ and its dependence of the depth 
on  $\eta$. One can 

1. measure the average depth  of variations and compare it with expectations;  

2. compare  the average depths for deep (e.g. $\eta = 0.58 - 1.20$) 
and  outer ($\eta =  1.20 - 1.57$) trajectories in the mantle;  

3. compare of  the average depths in the core and in the mantle.

Let us consider the first method. 
According to  Fig. \ref{fig:temp}
the depth $\langle D \rangle \propto \frac{1}{T}$ 
and $D(\eta)$ changes linearly with $\eta$ 
(see Fig.~\ref{Fig11}). Therefore  
we can write for the averaged depth  for all mantle trajectories   
\be
\langle D \rangle = \frac{2}{\pi} \left( 0.63 + 
0.19 \frac{T_c^{st}}{T_c} \right) \cdot 10^{-3}, 
\label{avdepth}
\ee
where $T_c^{st}$ is the standard value of temperature in the center of the Sun. 
The   first term in the brackets
corresponds to the half  depth of oscillations at the surface
(trajectories with large $\eta$),  the second one corresponds  
to the deepest trajectories in the mantle for $T_c = T_c^{st}$.
For $T_c =  T_c^{st}$ we obtain  from (\ref{avdepth}) 
$\langle D \rangle = 0.51  \cdot 10^{-3}$,  
whereas for $T_c =  0.5 T_c^{st}$: $\langle D \rangle = 0.64 \cdot 10^{-3}$. 
So,  decrease of the $T_c$ by factor 2 leads to 
increase of the average depth by $20\%$.  As we have found in the 
previous subsection the deviation of the average depth 
from 0 can be established with  significance about $2.5 \sigma$,  
so its $20\%$ change would correspond to about $0.5 \sigma$.

Consider the second method. According to Fig. \ref{Fig11}, 
the average depth   in the outer part of mantle ($\eta =  1.20 - 1.57$), 
equals  $\langle D \rangle = 0.64  \cdot 10^{-3}$, whereas in  the inner part  
($\eta = 0.58 - 1.20$): $\langle D \rangle = 0.33  \cdot 10^{-3}$.  
We take that the number of events in the inner and outer intervals of $\eta$ 
are the same and equal  $N_N/2$,  so that  variations in the 
outer and inner ranges of $\eta$ equal 
$6.7 \cdot 10^{3}$ and $3.4 \cdot 10^{3}$ correspondingly. 
The statistical error is  $3.9 \cdot 10^{3}$, therefore  variations can be established at 
$1.7 \sigma$ (outer) and $0.9 \sigma$ (inner). 
The ratio of the depths (which is sensitive to temperature)
equals 
$$
\frac{\langle D \rangle_{in}}{\langle D \rangle_{out}} = 0.51 \pm 0.66, 
$$ 
where the $1\sigma$ statistical errors is indicated.  
So,  even decrease of the ratio by factor 2 will correspond to $0.4 \sigma$. 

Sensitivity is low but this is independent measurement of the width. 
Clearly, factor of 2 uncertainty in central  temperature is excluded,  
e.g.,  by measurements of the boron neutrino flux.

%%%%%%%%%%%ffff6%%%%%%%%%%%%%%%%%%%%%%%%%%%
%%\begin{figure}[b]
%%\includegraphics[width=0.9\columnwidth, height=5.7cm]{SCT7.png}
%%\caption[...]{Dependance of  average  annual weight function on 
%%the nadir angle ($\eta$  is in radians) of neutrino trajectory
%%for Pyh\"asalmi (solid line), Kamioka (dotted line) and  Daya Bay-JUNO (dashed line) sites.
%%\label{Fig0}}
%%\end{figure}
%%%%%%%%%%%%%%%%%%%%%%%%%%%%%%%%%%%%%%%%%%%%%%%%

%%%%%%%%%%%%%%%%%%%%%%%%%%%%%%%%%%%%%%%%%%%%%%%%%%%%%%%%%%%%%%%%%%%
\subsection{Determination of  $\Delta m^2_{21}$}
%%%%%%%%%%%%%%%%%%%%%%%%%%%%%%%%%%%%%%%%%%%%%%%%%%%%%%%%%%

The oscillatory pattern of Fig.~\ref{Fig11}  
depends on precise value of $\Delta m^2_{21}$. 
Uncertainty in  $\Delta m^2_{21}$,   $\delta(\Delta m^2_{21})$,  would not 
influence the pattern if 
\be
\frac{\delta(\Delta m^2_{21})}{\Delta m^2_{21}} \ll \frac{l_\nu}{L_{max}} \approx 0.3\% . 
\ee
The present accuracy of determination of $\Delta m^2_{21}$ 
is $2.5 \%$  and future experiments (e.g. JUNO-reactors)
will not be able to reach $0.1 \%$ accuracy. 
So,  $\Delta m^2_{21}$ 
should be extracted from the $^7$Be-neutrino studies simultaneously 
with other measurements.

The mass splitting $\Delta m_{21}^2$ determines the depth, the average value 
of probability and the length of oscillations. Correspondingly,  there are
several  different methods to measure $\Delta m_{21}^2$.
The depth of variations of $A_N$ and its average
value according to Eqs. (\ref{osclength}) and (\ref{const})
are  inversely  proportional to the splitting:
$D  \propto 1/\Delta m_{21}^2$,
$A_e \propto 1/\Delta m_{21}^2$.
As we have established in the previous subsection,
$A_e$ and $D$  
can be distinguished from 0 at about
$2.3\sigma$ level. Consequently, these measurements
will determine $\Delta m_{21}^2$ with about $1/2.3 \approx 40\%$ accuracy
at $1 \sigma$ level. 

Since there are  many (up to 400,  according to Fig. \ref{Fig11})  
periods of oscillations, dependence of the oscillation period on $\Delta m_{21}^2$
can provide much better determination of $\Delta m_{21}^2$.
Indeed, the error $\delta(\Delta m_{21}^2) \sim (1/300) \Delta m_{21}^2$ 
would lead to substantial change of the oscillatory picture.
Clearly knowledge of exact values of $L$  for the detected events is crucial.

The analysis can be performed using, e.g., 
the Lomb-Scargle \cite{Lomb:1976wy},  
\cite{Scargle:1982bw} method which is valid for uneven time intervals.  
Let   
\be
N_N^j = N_D^j (1 + A_N^j)
\label{incc}
\ee 
be  the observed  number of events in the $j-$  bin with the average nadir angle $\eta_j$.  
This corresponds to the true value  of $\Delta m^2_{21} = \Delta m^{true ~2}_{21}$.  
The first term in (\ref{incc}) can be considered as fluctuating noise, 
whereas the second one as periodic signal to be extracted. 
For constant density  we can use expression  
%%$A_N^j = \xi \sin^2 \Delta_m L_j/2$,  $\xi \equiv - c_{13}^2 \epsilon f$ 
(\ref{const}), so that 
\be
N_N^j \approx  N_D^j (1 + \frac{1}{2}c_{13}^2 \epsilon f \cos \Delta_m^{true} L_j). 
\ee
The  method require that the mean value  of the background is zero. 
So,  we need to subtract the average day signal $\bar{N}_D$ from $N_N^j$.  
Then the Lomb-Scargle periodogram is defined as  
\be
P_{LS}= {1 \over {\it n}}{(\sum_j  [N_N^j (\Delta_m^{true}) - \bar{N}_D] \cos \Delta_m^{fit} x_j)^2  
\over  \sum_j \cos^2 \Delta_m^{fit} x_j },  
\label{lssum}
\ee
where $n$ is the total number of bins and $\Delta_m^{fit}$ corresponds to the fit value of 
$\Delta m^2_{21}$. 
The true value of  $\Delta m^2_{21}$ can be obtained by   
varying $P_{LS}$ over $\Delta m^{fit~2}_{21}$: it corresponds to 
maximum  of $P_{LS}$. 

For illustration we can perform  simplified computations using 
constant density profile and spherically symmetric Earth. In this case 
the $N_N$ is exactly periodic function of $\cos \eta$. 

The Lomb-Scargle periodogram is based on the discrete Fourier transform, but  
in the case of many bins of even size we 
can substitute summation in eq. (\ref{lssum}) 
by integration over $x \propto \cos \eta$ from 0 to $x_{max}$
(here $x$ is the length of the trajectory).  
So, 
\be
P_{LS} = \frac{
[ x_{max}^{-1} \int dx [N_N (x) - \bar{N}_D]\cos \Delta_m^{fit} x]^2}
{n x_{max}^{-1} \int dx \cos^2 \Delta_m^{fit}x }, 
\label{pls12}
\ee
where 
\be
N_N (x) \approx  N_D (x) \left(1 + \frac{1}{2}    
c_{13}^2 \epsilon f \cos \Delta_m^{true} x \right).
\ee
The integration in (\ref{pls12}) gives  
\be
P_{LS} =  N_D^2  \frac{(c_{13}^2 \epsilon f)^2}{8} \left|\frac{\sin \Delta \Phi}{\Delta \Phi}\right|^2,  
\label{lslsls1}
\ee
where 
\be
\Delta \Phi  = (\Delta_m^{true} - \Delta_m^{fit}) x_{max}.
\ee

Using  expression for $\Delta_m$ in vacuum we find the following relation 
for $\delta(\Delta m_{21}^2) \equiv
\Delta m_{21}^{true~2} - \Delta m_{21}^{fit~2}$:
\be
\frac{\delta(\Delta m_{21}^2)}{\Delta m_{21}^2}
= \frac{\Delta \Phi}{4 \pi} \frac{l_\nu}{x_{max}}.
\label{deltadelta}
\ee
For $\Delta \Phi = \pi/2$, which corresponds to
half of the height of the peak in (\ref{lslsls1}),
Eq. (\ref{deltadelta}) gives 
\be
\frac{\delta(\Delta m_{21}^2)}{\Delta m_{21}^2} =
4 \cdot 10^{-4}.
\ee
These estimations do not take into
account  effect of averaging over the $^7$Be-neutrino
spectrum.  So, realistic 
accuracy will be above 0.001.

\subsection{Tomography of the Earth}
%%%%%%%%%%%%%%%%%%%%%%%%%%%%%%%%%%%%%%%%%%%%%%%%%%%%%%%%%%%%%%%%%%%%%%%%%%%%%%%

As we discussed in Sec. 3, the oscillatory pattern  encodes information
on the shape and density profile of the Earth.
So,  in principle one can perform tomography of the Earth 
with spatial precision comparable with the oscillation length. 
Let us summarize some dependences:

- Sudden change of the depth of variations of $A_N$ at $\eta \sim 0.58$
marks trajectories which start to cross the core. 
The corresponding  $\eta$ gives the position of the density jump between 
the mantle and the core, the size of  
change of the depth of oscillations reflects the size of the jump. 

- Detailed oscillation pattern in the core region is sensitive
to the parameters of the inner core.

- Small density jumps in the mantle trajectory range  and
modulations of the high frequency oscillations
encode information about the mantle structure:
position of borders between different regions
(inner mantle, outer mantle, {\it etc.}) and sizes of density
jumps at their borders. 

- The oscillatory picture is sensitive to
the shape of the Earth, in particular, to its deviation from
sphericity.

- Distortion of the periodic oscillatory curve is sensitive to local structures 
(mountains, seas, oceans, oil layers, {\it etc.}) \cite{ara3}.

Detectors with $0.1\%$ accuracy of measurements  will see (resolve) core and mantle. 
Other features will have lower significance and appear as
systematic errors in these detectors.

\section{5. Searches for sterile neutrinos}
%%%%%%%%%%%%%%%%%%%%%%%%%%%%%%%%%%%%%%%%%%%%%%%%%%%%%%%%%%

The parameter $\epsilon$, and consequently, 
the Earth matter effect
are inversely proportional to the mass splitting $\Delta m^2_{21}$.
Therefore with decrease of $\Delta m^2_{21}$ (for the
same value of mixing) the variation $A_N$  
increases. In this connection let us consider
sterile neutrinos with very small mass splitting, 
$\Delta m^2_{10} \equiv m^2_{0} - m^2_{1}
\ll \Delta m_{21}^2$. Here $m_0$ is the mass of new state. 
The matter parameter  for the $\nu_e - \nu_s$ system, 
$\epsilon_s$,  is determined by the potential
$V_s \approx 0.5 V_e$. 
Then the resonance, $\epsilon_s = 1$,  is achieved for
\be
\Delta m^2_{10} = \Delta m^2_{10 R} =
0.9 \cdot 10^{-7}\, \,   {\rm eV}^2. 
\label{dms-ster}
\ee
(for the mantle densities). In the core the 
resonance enhancement will be at about 2 time bigger splitting. 

For definiteness we will consider mixing
of sterile neutrino in the mass state $\nu_1$ characterized by  the
angle $\theta_s$. In this case  we obtain for the difference of the night
and day signals
\be
\Delta P \approx |U_{e1}|^2
(|A_{01}|^2 -|A_{00}|^2)(P_{1e}^s -|U_{e1}|^2),
\label{dpster}
\ee
where $A_{01}$ and  $A_{00}$ are the amplitudes of
transitions between the matter eigenstates:
$\nu_{0m} \rightarrow \nu_{1m} = \nu_1$
and $\nu_{0m} \rightarrow \nu_{0m} = \nu_0$
when neutrinos propagate from the center  
to the surface of the Sun. 
In (\ref{dpster}) we have taken into account that in the Sun,  for small mass splitting
the neutrinos are produced far above $\nu_1 - \nu_0$ resonance
and in initial state $\nu_{0m} \approx \nu_1$
which gives the factor $|U_{e1}|^2$ in Eq.
(\ref{dpster}).  Eq. (\ref{dpster}) is analogy of expression (\ref{probnue}).

Even for splitting (\ref{dms-ster})
the coherence will be lost due to separation of the
wave packets of the mass states on the way to the Earth.
So,  as in the active neutrino case,
independent fluxes of $\nu_1$  and  $\nu_0$
will arrive at the Earth surface. 

For oscillations in the Earth
we consider the $2\nu-$system $\nu_1 - \nu_0$ and
denote by $P_{1e}^s$ the probability of $\nu_1 \rightarrow \nu_e$ transition.
For estimations we use  the constant density
profile for the trajectories inside the Earth
with average value of density.
This is justified since due to smallness of  
$\Delta m^2_{21}$  the oscillation length is
much larger than the size of the Earth.
In this case
\be
P_{1e}^s -|U_{e1}|^2 = - \epsilon_s \sin^2 2\theta_{s}^m
\sin^2 \left(\frac{1}{2} \Delta_m L \right) ,
\label{const-s}
\ee
For relatively large mixing angle the adiabatic transition
$\nu_{0m} \rightarrow \nu_{0}$ inside the Sun
would lead to substantial additional
suppression of the $\nu_e$  $^7$Be-neutrino flux.
Theoretical accuracy of the flux is about $1.4\%$. 
Experimental accuracy (presently $5\%$) will be much better.
So, the survival probability should be larger than 0.98,
that is, the adiabaticity should be strongly broken with
$|A_{01}|^2 > 0.98$. Using results for active neutrinos
(see e.g. \cite{msw-rew}) we estimate that
this can be achieved if  $\sin^2 2\theta_{s} < 0.01$ for
splitting in (\ref{dms-ster}). 
Taking $|A_{01}|^2 - |A_{00}|^2 \approx 0.96$,  $|U_{e1}|^2 \approx 0.67$
and $P_D \approx 5/9$, we obtain from Eqs. (\ref{dpster}) and (\ref{const-s}) 
\be 
A_N  \approx  - 1.15 \epsilon_s \sin^2 2\theta_{s}^m 
\sin^2 \left(\frac{1}{2} \Delta_m L\right).
\label{dpster1}
\ee

Let us consider maximal allowed value
$\sin^2 2\theta_{s} = 0.01$.
In this case the width of the MSW resonance, 
$\Delta E  = 2 \tan 2\theta_{s} E \approx 0.2 E = 190$ kev, 
is much larger than the width of the line.
%%So,  the effect is determined by value of the probability at
%%the energy  of the $^7$Be line.
Changing $\Delta m^2_{10}$ within $(20 - 40) \%$ would put the line
at different points of the MSW resonance peak. 

In the resonance we have $\epsilon_s \approx \sin^2 2\theta_{s m}
\approx 1$ and therefore 
\be
A_e^{res} \approx  - 1.15 \sin^2 \frac{1}{2}\Delta_m L(\eta).  
\label{asym-s}
\ee
For the resonance value of $\Delta m^2_{21}$  (\ref{dms-ster}) 
the oscillation length in vacuum equals $l_{\nu} = 2.5 \cdot 10^4$ km
and in matter (in resonance) $l_m = l_{\nu}/\sin 2\theta_s
= 2.5 \cdot 10^5$ km, {\it i.e.} much larger than the diameter of the Earth.
For the deepest trajectory in the mantle, $L \approx 10^4$ km,
the phase equals $0.126$ rad.,  and according to Eq. (\ref{asym-s})
$A_e = 0.017$. This is an order of magnitude larger than
the effect for active neutrinos and of the order of the present accuracy.
For $\cos \eta = 0.4$ (middle trajectory in the mantle)
we find $A_e = 0.0045$. The effect monotonously increases as
$\propto 1/(\cos \eta)^2$, reaching maximum for the deepest trajectory
(middle of the night) and then it decreases down to zero.

For the core crossing trajectories the effect can be more complicated. 
The MSW resonance in the core 
leads to appearance of another peak, and the $^7$Be neutrino line can be in one 
peak or another.
Also the interplay of the effects in the mantle and the core may occur.

If $\Delta m^2_{10}$ is more than $40 \%$ larger than the resonance value,
the values of oscillation parameters become close to vacuum
values: $l_m \approx l_\nu$,  $\sin^2 2\theta_{s m} \approx 0.01$,
so that 
\be
A_e \approx  - 0.015 \epsilon_s \frac{\sin^2 2\theta_{s}}{10^{-2}}
\sin^2 \frac{1}{2} \Delta_m L(\eta).
\label{dpster1}
\ee
In this case still $\epsilon_s = O(1)$, 
and the oscillatory factor can be
of the order 1. Thus, the effect for the deepest trajectories in the mantle may  reach
$1\%$. Since the oscillation length becomes comparable with
the size of the Earth, the time dependence of the effect is 
more complicated, e.g. with two maxima symmetrically shifted from
the middle of the night.

With further increase of $\Delta m^2_{10}$ (outside the resonance region)
the size of the effect  decreases as $1/\Delta m^2_{10}$, 
and time dependence will acquire  an oscillatory form 
with increasing number of periods. 

Let us make similar estimation for
$\sin^2 2\theta_{s} = 0.001$.
In resonance the oscillation length equals 
$l_m = 8.5 \cdot 10^5$ km
(about 3 times bigger than in the previous case).
Correspondingly, the phases will be 3 times smaller, and 
the oscillatory factor will be an order of magnitude smaller 
As a result,  for the deepest mantle trajectory 
$A_e \sim 10^{-3}$.
Outside the resonance peak we obtain $A_e < 0.001$. 
Thus, the effect decreases as $\sin^2 2\theta_{s}$ everywhere.

%%%%%%%%%%%%%%%%%%%%%%%%%%%%%%%%%%%%%%%%%%%%%%%%%%%%%%%%%%%%%%%%%
\section{6. Conclusions}
%%%%%%%%%%%%%%%%%%%%%%%%%%%%%%%%%%%%%%%%%%%%%%%%%%%%%%%%%%%%%%%%%%%%%%%%%%%%%%%%%%%%%%%%%%

We explored in detail effects of propagation of
the solar $^7$Be neutrinos in the matter of the Earth.
We estimated a possibility to detect these
effects with  future large scintillator (or scintillator uploaded)
detectors which will have $0.1\%$ accuracy of measurements.

The main features of the propagation are determined by
low energy of the $^7$Be neutrinos and their narrow
energy spectrum.
Oscillations in the Earth are pure matter effect and they are related to
transitions between the mass eigenstates.

Physics of oscillations is determined by two  accidental coincidences.
Due to low energies, the oscillation length, $\approx 30$ km,
is of the order of small structures of the Earth profile and non- sphericity 
of the Earth.
The width of the spectrum is comparable with the period of the oscillatory curve in
the energy scale. So that depending on length of trajectory
(nadir angle) one should observe different degree of
averaging. In configuration space this is equivalent to
partial loss of coherence due to shift of the wave packets
of different eigenstates of the Hamiltonian in the course of propagation inside the Earth.
The size of the packets due to spread on the way from
the Sun is several orders of magnitude larger than the shift. 

The main observable is the oscillatory variation
(mainly suppression) of the
signal in time (with  nadir angle) during the night.
The depth of oscillations  changes with time since 
the averaging of oscillations  becomes stronger with increase
of the length of trajectory. The depth is the largest
for shallow trajectories and it is the smallest for the
deepest mantle trajectories. It sharply increases when
trajectory crosses the core. Small density jumps in the mantle and the core 
produce modulations of the oscillatory curve. 

We find that for illustrative configuration of experiment
(100 kt, 5 years of exposure) the Earth matter effect
can be established  at $2. 3  \sigma$ level; 
the width of the $^7$Be-neutrino line can be determined  with factor of 2 
accuracy at  $0.5\sigma$; $\Delta m^2_{21}$ can be measured  with
accuracy $0.1\%$. The presence of the core
of the Earth and its border can be seen by the detector 
(close to equator) at $2 \sigma$.
The sensitivity  can be enhanced if one uses larger exposure time 
(e.g. 10 years) or larger fiducial 
volume of the detector.

Determination of other  characteristics:
size of region in the Sun where the Be neutrinos are produced,
detailed tomography of the Earth (small scale structures,
layers in the mantle and the core)  will require further
substantial increase of the detector size.

At this level of accuracy it is not possible to measure whole oscillatory 
curve which contains interesting information about structure of the Earth. 
That would open up a possibility  to use the Sun as the scanner of the Earth 
to perform tomography. However, estimations show that the required sensitivity is 
not by several orders of magnitude higher. 
%%So the issue is    not completely out of question. 

One can perform searches of sterile neutrinos 
with mixing  $\sin^2 2 \theta_s = 10^{-3} - 10^{-2}$    
in wide range of $\Delta m^2_{10} > 10^{-8}$ eV$^2$, 
especially in the resonance
region of mass splitting around $10^{-7}$ eV$^2$.

%%Future detectors with sensitivity to $^7$Be neutrino signal 
%%at the level  $10^{-3}$ can establish  
%%varios matter effects at the  about $(2 - 3) \sigma$ level.

\section*{Acknowledgements}

Ara Ioannisiyan  thanks ICTP for visit in 2014, when part of this work has been accomplished.

%%%%%%%%%%%%%%%%%%%%%%%%%%%%%%%%%%%%%%%%%%%%%%%%%%%%%%%%%%%%%
%\bibliography{apssamp}

\end{document}